\documentclass[12pt]{article}
\usepackage{amssymb,amsmath}
\usepackage{cite}
\usepackage{epsfig}
\usepackage{graphics}

\newlength{\dinwidth}
\newlength{\dinmargin}
\newtheorem{theorem}{Theorem}[section]

\newtheorem{lemma}[theorem]{Lemma}

\newtheorem{definition}[theorem]{Definition}

\newcommand{\ie}{{\it i.e.\ }}

\def\idty{{\leavevmode\hbox{\rm 1\kern -.3em I}}}

\def\idty{{\leavevmode\hbox{\rm 1\kern -.3em I}}}

\def\RR{{\mathbb R}}

\def\NN{{\mathbb N}}

\newcommand{\cf}{\textit{cf.\ }}

\def\rest{\upharpoonright}

\def\cD{{\cal D}}

\def\cH{{\cal H}}

\def\cL{{\cal L}}

\def\cP{{\cal P}}

\def\cR{{\cal R}}
\def\cS{{\cal S}}

\def\cW{{\cal W}}

\def\NN{{\mathbb N}}

\def\RR{{\mathbb R}}

\newcommand{\bcW}{{\mbox{\boldmath$\cal W$}}}

\newcommand{\bcp}{{\mbox{\boldmath$p$}}}
\newcommand{\bcq}{{\mbox{\boldmath$q$}}}
\newcommand{\sbcp}{{\mbox{\scriptsize \boldmath$p$}}}

\begin{document}

\title{Warped Convolutions: A Novel Tool in the \\
Construction of Quantum Field Theories\footnote{Talk given
by DB at Ringberg Symposium in honor of W.\ Zimmermann, 
February 2008}} 

\author{{\Large Detlev Buchholz\,$^a$ \ and \  
Stephen J.\ Summers\,$^b$ }\\[5mm]
${}^a$ Institut f\"ur Theoretische Physik, 
Universit\"at G\"ottingen, \\ 37077 G\"ottingen, Germany  \\[2mm]
${}^b$ Department of Mathematics, 
University of Florida, \\ Gainesville FL 32611, USA}

\date{June 2, 2008}

\maketitle 

\begin{abstract}  \noindent
Recently, Grosse and Lechner introduced a novel deformation procedure 
for non--interac\-ting quantum field theories, giving rise to
interesting examples of wedge--localized quantum fields with a 
non--trivial scattering matrix. In the present article  
we outline an extension of this procedure to the general framework of 
quantum field theory by introducing the concept of warped 
convolutions: given a theory, this construction provides   
wedge--localized operators which commute at spacelike 
distances, transform covariantly under the underlying representation 
of the Poincar\'e group and admit a scattering theory. The 
corresponding scattering matrix is nontrivial but breaks the Lorentz 
symmetry, in spite of the covariance and wedge--locality properties 
of the deformed operators.
\end{abstract}

\section{Introduction}

     Recent advances in algebraic quantum field theory have led to
purely algebraic constructions of quantum field models on Minkowski
space, both classical and noncommutative
\cite{BrGuLo2,Sch,Le,BuLe,Le2,MuSchYng,BuSuads,BuSu2,Le3,GrLe}, many
of which cannot be constructed by the standard methods of constructive
quantum field theory. Some of these models are local and free, some
are local and have nontrivial S-matrices, and yet others manifest only
certain remnants of locality, though these remnants suffice to enable
the computation of nontrivial two--particle S-matrix elements.

     In a recent paper \cite{GrLe}, Grosse and Lechner have presented
an infinite family of quantum fields which, taken as a whole, are
wedge--local and Poincar\'e covariant and which have nontrivial
scattering.  They produce this family by deforming the free quantum
field in a certain manner, motivated by the desire to understand the
field as being defined on noncommutative Minkowski space. However, as
they point out, one can forget the original motivation and view the
resulting deformed fields as being defined on classical Minkowski
space. It is, however, essential to the arguments of \cite{GrLe} that
the {\it free} field is deformed.

     In this paper we present a generalization of their deformation
which can be applied to any Minkowski space quantum field theory in
any number of dimensions. This deformation results in a one parameter
family of distinct field algebras which are wedge--local and covariant
under the representation of the Poincar\'e group associated with the
initial, undeformed theory. It turns out that also the $S$--matrix
changes under this deformation, and the deformed $S$--matrix breaks
the Lorentz symmetry, in spite of the Lorentz covariance of the
deformed theory.  When taking the free quantum field as the initial
model, our deformation coincides with that of Grosse and Lechner.

     The deformation in question involves an apparently novel
operator--valued integral, whose mathematical definition requires 
some care. Apart from the operators which are to be integrated, 
it involves a unitary representation of the additive group
$\RR^d$, $d \geq 2$, satisfying certain properties which arise
naturally when considering relativistic quantum field theories on two
(or higher) spacetime dimensional Minkowski space. 
We outline in Sec.\ 2 the intriguing properties of this 
integral; proofs will be given elsewhere. In Sec.\ 3
we apply these results to quantum field theories to obtain the results
mentioned above. Finally, in Sec.\ 4 we indicate some paths of 
further investigation suggested by these results.

\section{Warped convolutions}

     In order to draw attention to what may be regarded as the
mathematical core of the deformation studied in this paper, we
consider a quite general setting which covers both the case of
Wightman Quantum Field Theory considered in \cite{GrLe} and the case
of Algebraic Quantum Field Theory \cite{Haag}.

     We shall assume the existence of a strongly continuous unitary
representation $U$ of the additive group $\RR^d$, $d \geq 2$, on some
separable Hilbert space $\cH$.  The joint spectrum of the generators
$P$ of $U$ is denoted by $\mbox{sp} \, U$ and will be further
specified in the following section. Let $\cD$ be the dense subspace of
vectors in $\cH$ which transform smoothly under the action of $U$, \cf
\cite{FrHe}. We consider the set ${\mathfrak F}$ of all operators $F$
which have $\cD$ in their domain of definition and are smooth under
the adjoint action $\alpha_x(F) \doteq U(x) F U(x)^{-1}$ of $U$ in the
following sense: for each $F \in {\mathfrak F}$ there is some 
$n \in \NN$ such that the operator valued function 
$x \mapsto (1 + |P|^2)^{-n} \alpha_x(F) (1 + |P|^2)^{-n} $ is arbitrarily 
often differentiable in norm, where $|P|^2$ denotes the sum of the squares
of the generators of $U$.  It is easily seen that ${\mathfrak F}$ is a
unital *--algebra.

     Within this framework one can establish a deformation procedure
for the elements of ${\mathfrak F}$. The basic ingredient in this
construction is the spectral resolution $E$ of the unitary group $U$,
$$   U(x) = e^{iPx} =
\int e^{ipx} \, dE(p) \, , \quad x \in \RR^d \, ,$$
where the inner product on $\RR^d$ is arbitrary here and will be
fixed later. Given any skew--symmetric $d \times d$--matrix $Q$, 
\ie $q \, Q p = - p \, Q q$ for $p,q \in \RR^d$, one can  
give meaning to the operator valued integrals for any $F \in {\mathfrak F}$
\begin{equation} \label{intdef}
{}_QF \doteq   \int \! \alpha_{Qp}(F) \, dE(p) 
\, , \quad F_Q \doteq \int \! dE(p)  \, \alpha_{Qp}(F) \, . 
\end{equation}
These left and right integrals are defined on the domain $\cD$ in the sense
of distributions. Moreover, the resulting operators are smooth
with regard to the adjoint action of $U$ in the sense 
explained above; hence ${}_QF,  F_Q  \in {\mathfrak F}$. 
We omit the proof and only note that the above integrals 
may be viewed as warped (by the matrix $Q$) convolutions 
of $F$ with the spectral measure $dE$.  

The above integrals have a number of remarkable properties,
which are crucial for their application to quantum field 
theory. We begin by noting the at first sight surprising fact
that the left and right integrals coincide.

\begin{lemma} \label{leftright} 
Let $F \in {\mathfrak F}$. Then ${}_QF = F_Q$.
\end{lemma}

     The proof of this lemma requires the proper treatment of 
expressions such as $dE(p) F dE(q)$ (which are not  
product measures) as well as the discussion of subtle domain
problems. We therefore forego here a rigorous argument.
Yet, in order to display the significance of the 
skew symmetry of the matrix $Q$ for the result,
we indicate the various steps in the proof. Making use several times of 
the relation $dE(p) f(P) = dE(p) f(p) = f(P) dE(p)$, which holds for any 
continuous function $f$, we get the following 
chain of equalities, which are justified below.
\begin{equation*}
\begin{split}
& F_Q = \int \! dE(p) \, \alpha_{Qp}(F) \\ 
& = \int \! dE(p) \, U(Qp) F U(Qp)^{-1} \! \int dE(q) \\
& = \iint \! dE(p) \,  U(Qp) F \, U(Qp)^{-1} \, dE(q) \\
& = \iint \! dE(p) \, F e^{ipQq} \, dE(q)  \\
& = \iint \! dE(p) \,  e^{ipQq} F \, dE(q) \\
& = \iint \! dE(p) \, U(Qq) F U(Qq)^{-1} \, dE(q) \\
& = \int  \! \alpha_{Qq}(F) \, dE(q) = {}_Q F \, .
\end{split}
\end{equation*}
In the second equality we made use of $\int dE(q) = 1$,  
in the third one we relied on the fact that the preceding 
expression can be rewritten as a double integral, and in
the fourth one we used the skew symmetry of $Q$, 
implying $dE(p) \, e^{- iPQp} =  dE(p) $
and $e^{- iPQp} \, dE(q) =  e^{- iqQp} dE(q) = e^{ipQq} dE(q)$.
The fifth equality then follows, since $e^{ipQq}$ is a 
c--number, and the sixth one is a consequence of
$dE(p)  \, e^{ipQq} = dE(p) \,  e^{iPQq}$
and $dE(q) =  e^{- iPQq} \, dE(q) $. In the 
last step we made use once again of $\int dE(p) = 1$. 

     It can be inferred from the defining relations (\ref{intdef})
that $({}_Q F)^{\, *} \supset {F^*}_Q$. Thus, as an immediate
consequence of the preceding lemma, one finds that the operation of
taking adjoints commutes with the warped convolution in the following
sense.

\begin{lemma} \label{adjoint}
Let $F \in {\mathfrak F}$. Then ${F_Q}^{\, *} \supset {F^*}_Q$.
\end{lemma}

     It is also noteworthy that $(F_{Q_1})_{Q_2} = F_{Q_1 + Q_2}$, for
any $F \in {\mathfrak F}$ and skew symmetric matrices $Q_1, Q_2$.  In
the next lemma we exhibit commutation properties of certain specific
elements of ${\mathfrak F}$, which are preserved by the deformation
procedure. The shape of the spectrum $\mbox{sp} \, U$ of the unitary
group $U$, which coincides with the support of the spectral measure
$dE$, enters in the formulation of this result.

\begin{lemma} \label{commutation}
Let $F, G \in {\mathfrak F}$ be such that 
$$\alpha_{Qp}(F) \, \alpha_{-Qq}(G) = 
\alpha_{-Qq}(G) \, \alpha_{Qp}(F) $$ for all  $p,q \in 
\mbox{\rm sp} \, U$. Then,  
$$F_Q \, G_{-Q} = G_{-Q} \, F_Q \, . $$ 

\end{lemma}

     Again, the rigorous proof of this result is plagued by
technicalities and will not be given here. But the following formal
steps, which are explained below, display the basic facts underlying
the argument.
\begin{equation*}
\begin{split}
& F_Q \, G_{-Q} = 
\int \! dE(p) \, \alpha_{Qp}(F) \int \! dE(q) \, \alpha_{-Qq}(G) \\
& = \int  \! dE(p)  \, \alpha_{Qp}(F) \int \alpha_{-Qq}(G)  \,  dE(q) \\
& = \iint dE(p)  \, \alpha_{Qp}(F)  \, \alpha_{-Qq}(G)  \, dE(q) \\
& = \iint dE(p)  \, \alpha_{-Qq}(G)  \, \alpha_{Qp}(F)  \, dE(q) \\
& = \iint dE(p)  \, U(-Qq) G U(-Qq)^{-1}  \, U(Qp) F   U(Qp)^{-1}  \, dE(q) \\
& = \iint dE(p)  \, e^{-ipQq}  G U(Qq + Qp) F e^{-iqQp}   \, dE(q) \\
& = \iint dE(p)  \, U(- Qp) G U(Qp + Qq) F U(- Qq)   \, dE(q) \\
& = \iint  dE(p)  \, \alpha_{-Qp}(G)  \, \alpha_{Qq}(F)  \, dE(q) \\
& = \int \! dE(p) \, \alpha_{-Qp}(G)  \, \int \! dE(q) \, \alpha_{Qq}(F) 
=  G_{-Q} \, F_Q \, .
\end{split}
\end{equation*}

\noindent In the second equality use was made of Lemma \ref{leftright}, 
the third equality
relies on the fact that the preceding product of operators 
can be presented as a double integral, and in the fourth equality
the commutation properties 
of the operators $F, G$ were exploited. The adjoint
action of $U$ is written out explicitly in the
fifth equality, and in the sixth equality the group law for $U$
as well as the relations $ dE(p)  \, e^{-iPQq} = dE(p)  \, e^{-ipQq} $
and $  e^{-iPQp}  \, dE(q) = e^{-iqQp}   \, dE(q) $ were used. 
The step to the seventh equality is accomplished by noting that
the phase factors in the preceding expression cancel in view
of the skew symmetry of $Q$, which also implies  
$dE(p) =  dE(p)  \,  e^{-iPQp}  $, $dE(q) =   e^{-iPQq} \, dE(q)$.
In the eighth equality the various unitaries are recombined into
the form of adjoint actions, and in the subsequent equality the double
integral is reexpressed as a product of simple integrals;  
Lemma \ref{leftright} is then used once again.

     We conclude this discussion of the warped convolution with 
a remark on its covariance properties. Let $\cL$ be a 
matrix group acting isometrically (with regard to the chosen inner product) 
on $\RR^d$ and let $\cP = \cL \ltimes \RR^d$
be the semidirect product of the two groups. We assume that
the unitary representation $U$ of  $\RR^d$ can be extended to 
a representation of $\cP$, denoted by the same symbol. Denoting
the elements of $\cP$ by $\lambda = (\Lambda, x)$, one then 
has $U(\lambda) U(y) = U(\Lambda y) U(\lambda)$ and consequently
$U(\lambda) dE(p)  = dE(\Lambda p) U(\lambda)$.
It follows from standard arguments that ${\mathfrak F}$ is stable 
under the action of $\cP$ given by $\alpha_\lambda (F) = U(\lambda) F
U(\lambda)^{-1}$. Moreover, 
\begin{equation*}
\begin{split} 
 U(\lambda) \Big( \int \! \alpha_{Qp}(F) \, dE(p) \Big)
U(\lambda)^{-1} 
& = \int \alpha_{\Lambda Q p}(U(\lambda) F U(\lambda)^{-1})  \, dE(\Lambda p)  \\
& = \int \alpha_{\Lambda Q \Lambda^{-1} p}  \, (U(\lambda) F
U(\lambda)^{-1})  \, dE(p)  \, .
\end{split}
\end{equation*}
Note that the matrix $\Lambda Q \Lambda^{-1}$ is again skew 
symmetric with regard to the chosen inner product. 
We state the above result in the form of a lemma for later reference.

\begin{lemma} \label{covariance}
Let $F \in {\mathfrak F}$, let
$Q$ be any skew symmetric matrix and let 
$\lambda = (\Lambda, x)$ be any element of $\cP$. Then 
$$\alpha_\lambda (F_Q) = \big(\alpha_\lambda(F)\big){}_{\Lambda Q
  \Lambda^{-1}} \, .$$
\end{lemma}

     With these results we have laid the foundation for the application 
of warped convolutions to quantum field theory.

\section{Deformations of quantum field theories}

     We turn now to the discussion of local quantum field theories in
Minkowski space and their deformations. Identifying $d$--dimensional
Minkowski space with the manifold $\RR^d$, the Lorentz inner product
is given in proper coordinates by 
$x y = x_0 y_0 - \sum_{i=1}^{d-1} x_i y_i $. 
Any given quantum field theory on $\RR^d$ may then be
described as follows: there is a continuous unitary representation $U$
of the Poincar\'e group $\cP = \cL \ltimes \RR^d$ on a separable
Hilbert space $\cH$, where $\cL$ is the identity component of the
group of Lorentz transformations and $\RR^d$ the group of spacetime
translations. The joint spectrum of the generators $P$ of the abelian
subgroup $U \rest \RR^d$ is contained in the closed forward lightcone
$V_+ = \{ p \in \RR^d : p_0 \geq |\bcp|\}$ and there is a, up to a
phase unique, unit vector $\Omega \in \cH$, representing the vacuum,
which is invariant under the action of $U$.

     We assume that the underlying local field operators and
observables generate a unital *--algebra ${\mathfrak A} \subset
{\mathfrak F}$, where ${\mathfrak F}$ is the algebra of smooth
operators with respect to the translations $U \rest \RR^d$ introduced
in the preceding section. In the Wightman setting of quantum field
theory this assumption obtains if the underlying fields satisfy
polynomial energy bounds \cite{FrHe}.  In the framework of local
quantum physics, where one deals with von Neumann algebras of bounded
operators, one has to proceed to weakly dense subalgebras of elements
smooth with respect to the action of the translation subgroup. So in
both settings this assumption does not impose any significant
restriction of generality and covers all models of interest.

     The detailed structure of the theory is of no relevance
here. What matters, however, is the assumption that one can identify 
all fields and observables which are localized in certain specific
wedge--shaped regions, called wedges, for
short. We fix a standard wedge (see Figure 1)
$$\cW_0 \doteq \{ x \in \RR^d : x_1 \geq |x_0| \}$$ 
and note that in $d > 2$ dimensions all other wedges $\cW$ can
be obtained from $\cW_0$ by suitable Poincar\'e transformations,
$\cW = \lambda \cW_0$, $\lambda \in \cP$. In $d=2$ dimensions
this statement only holds true if one also includes the spacetime reflections
in $\cP$. 

\vspace*{5mm}
\begin{figure}[h]
\begin{center}
\hspace*{15mm} \epsfig{file=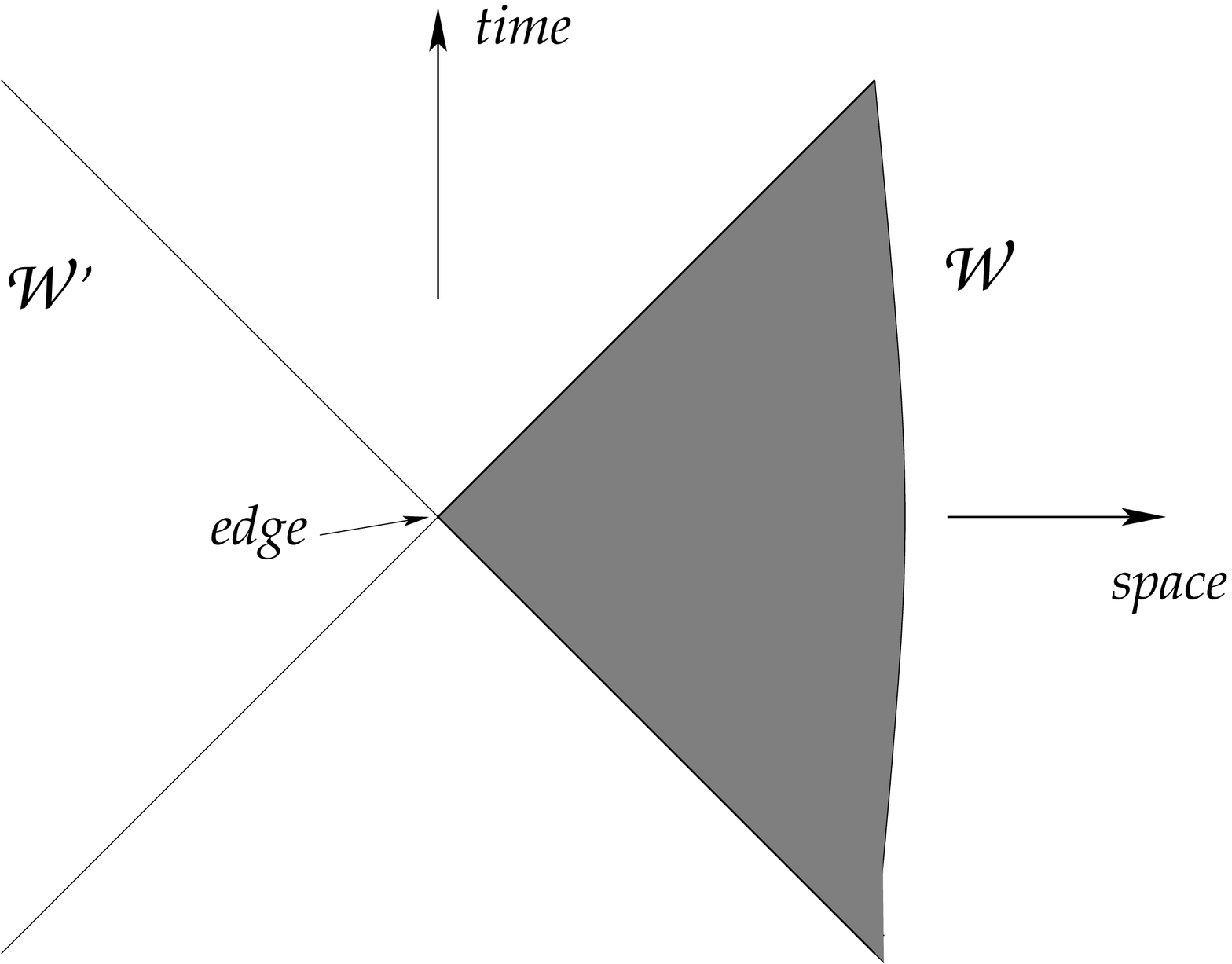,width=3.0in}
\end{center}
\caption{A wedge $\cW$, its causal complement $\cW^{\, \prime}$ and their
  common edge} 
\epsfxsize80mm
\end{figure}

\noindent Denoting by $\bcW = \{ \cW \subset \RR^d \}$
the set of all wedges in $\RR^d$, we consider for any given $\cW \in \bcW$ the 
*--algebra ${\mathfrak A}(W) \subset {\mathfrak A}$ generated by
all fields and observables localized in $\cW$. We call the algebras
${\mathfrak A}(W)$ wedge--algebras. It is apparent 
from the definition that ${\mathfrak A}(W_1) \subset {\mathfrak  A}(W_2)$ 
whenever $\cW_1 \subset \cW_2$, \ie isotony holds. The covariance, locality 
and Reeh--Schlieder property of the underlying theory can then be
expressed in terms of the wedge algebras as follows:
\begin{enumerate}
\item[(a)] \textit{Covariance:} \ $\alpha_\lambda ({\mathfrak A}(\cW)) =
U(\lambda) {\mathfrak A}(\cW) U(\lambda)^{-1}           
= {\mathfrak A}(\lambda \cW)$ for all $\cW \in \bcW$ and $\lambda \in \cP$. 
\item[(b)] \textit{Locality:} \ ${\mathfrak A}(\cW^{\, \prime})  
\subset {\mathfrak A}(\cW)^\prime $, $\cW \in
  \bcW$, where $\cW^\prime$ denotes the closure of the 
causal complement of $\cW$ and
${\mathfrak A}(\cW)^\prime$ the relative commutant of 
${\mathfrak A}(\cW)$ in ${\mathfrak F}$.
\item[(c)] \textit{Reeh--Schlieder property:} \ 
$\Omega$ is cyclic for any ${\mathfrak A}(\cW)$, $\cW \in \bcW$. 
\end{enumerate}

     We mention as an aside that these assumptions also cover   
quantum field theories on non--commutative Minkowski space
(Moyal space), as considered for example in \cite{GrLe}. These spaces
are described by non--commuting coordinates $X_\mu, X_\nu$
satisfying the commutation relations
$$ [X_\mu , X_\nu ] = i \, \theta_{\mu \nu} 1 \, , $$
where $\theta_{\mu \nu} = - \theta_{\nu \mu}$ are real constants. 
If  the dimension of the spacetime satisfies $d > 2$,
there exist lightlike coordinates $X_\pm$ with
$[X_+,X_-] = 0$ which can thus be simultaneously diagonalized. 
Hence fields and observables on such spaces can  
be localized in wedges $\cW$, yet they are 
dislocalized along the directions of the edges of these wedges. 
The wedge algebras are in general sufficient to reconstruct
the algebras corresponding to arbitrary causally closed 
regions $\cR$. These are given by 
$$ {\mathfrak A}({\cR}) \doteq \bigcap_{\cW \supset \cR} \, 
{\mathfrak A}(\cW) $$
and inherit from the wedge algebras both locality and covariance 
properties.  Yet in theories on non--commutative
Minkowski space, where fields and observables cannot be localized 
in bounded regions, the corresponding algebras are trivial  
and consequently do not manifest the Reeh--Schlieder property. 

     Given a theory as described above, we can now apply the
deformation procedure established in the preceding section.
To this end, we fix the standard wedge $\cW_0$ 
and pick a corresponding $d \times d$--matrix $Q_\kappa$, which with
respect to the chosen proper coordinates has the form
$$Q_\kappa \doteq 
\left( \begin{array}{ccccc}  0 & \kappa & 0 & \cdots & 0 \\ 
                       \kappa &    0   & 0 & \cdots & 0 \\
                            0 &    0   & 0 & \cdots & 0 \\    
                       \vdots & \vdots & \vdots & \ddots &  \vdots \\
                            0 &    0   & 0 & \cdots & 0 \end{array} \right) $$
for some fixed $\kappa > 0$. Note that this matrix is skew symmetric
with respect to the Lorentz inner product. The following basic 
facts pointed out in \cite{GrLe} are crucial for the subsequent 
construction.
\begin{enumerate}
\item[(i)] 
Let $\lambda = (\Lambda,x) \in \cP$ 
be such that $\lambda \cW_0 \subset
\cW_0 $. Then $\Lambda Q_\kappa \Lambda^{-1} = Q_\kappa$. 
\item[(ii)] 
Let $\lambda^\prime = (\Lambda^\prime, x^\prime) \! \in \! \cP$ 
be such that $\lambda^\prime \, \cW_0 \! \subset
\! \cW_0{}^\prime $. Then $\Lambda^\prime Q_\kappa \Lambda^{\prime \, -1} \!
= \! - Q_\kappa$.
\item[(iii)] $Q_\kappa  V_+ = \cW_0$.
\end{enumerate}
It is an immediate consequence of (i) that for
any two Poincar\'e transformations $\lambda_i = (\Lambda_i,x_i)$, $i=1,2$,
such that $\lambda_1 \cW_0 = \lambda_2 \cW_0$, one has
$\Lambda_1 Q_\kappa \Lambda_1^{-1} = \Lambda_2 Q_\kappa
\Lambda_2^{-1}$. Indeed, $\lambda_2^{-1} \lambda_1 =
\big(\Lambda_2^{-1} \Lambda_1, \Lambda_2^{-1}(x_1 - x_2) \big)$ 
maps $\cW_0$ onto itself, hence 
$\Lambda_2^{-1} \Lambda_1 Q_\kappa \Lambda_1^{-1} \Lambda_2 = Q_\kappa$.

     After these preparations we can now proceed from the 
given family of wedge algebras to a new ``deformed'' family 
with the help of the warped convolutions introduced in the preceding
section. For $\cW \in \bcW$ the corresponding 
deformed algebras ${\mathfrak A}_\kappa(\cW)$ are   
defined as follows.

\begin{definition} \label{def}
Let $\cW \in \bcW$ and let $\lambda = (\Lambda, x) \in \cP$ be such 
that $\cW = \lambda \cW_0$. The associated  algebra 
${\mathfrak A}_\kappa(\cW)$ is 
the polynomial algebra generated by all warped operators 
$A_{\Lambda Q_\kappa \Lambda^{-1}}$ with $A \in {\mathfrak A}(\cW)$. 
\end{definition}

\noindent 
Note that according to the preceding remarks this definition is consistent,
since it does not depend on the particular choice of the Poincar\'e
transformation $\lambda$ mapping $\cW_0$ onto $\cW$.
Moreover, by Lemma \ref{adjoint}, each ${\mathfrak A}_\kappa(\cW)$
is a *--algebra. 
We will show that the algebras ${\mathfrak A}_\kappa(\cW)$ have all 
desired properties of wedge algebras in a quantum field theory. 

     The isotony of the algebras ${\mathfrak A}_\kappa(\cW)$
is a consequence of the fact that if $\cW_1 \subset \cW_2$, 
these wedges can be mapped onto each other by a pure translation.
Hence there are Poincar\'e transformations $\lambda_i = (\Lambda,
x_i)$, $i = 1,2$, with the same $\Lambda$ mapping $\cW_0$
onto $\cW_1$ and $\cW_2$, respectively. As  
${\mathfrak A}_\kappa(\cW_1)$, ${\mathfrak A}_\kappa(\cW_2)$ 
are generated by the operators $A_{\Lambda Q_\kappa \Lambda^{-1}}$ 
with $A \in {\mathfrak A}(\cW_1)$ and 
$A \in {\mathfrak A}(\cW_2)$, respectively, the isotony
of the original wedge algebras implies 
${\mathfrak A}_\kappa(\cW_1) \subset {\mathfrak A}_\kappa(\cW_2)$
whenever $\cW_1 \subset \cW_2$. 

     For the proof of covariance we make use of
Lemma~\ref{covariance}. Let $\cW = \lambda_\cW \cW_0$ with
$\lambda_\cW =(\Lambda_\cW, x_\cW)$ and let $\lambda = (\Lambda, x)$. 
As the original theory is covariant, one has 
$\alpha_{\lambda_\cW}({\mathfrak A}(\cW_0)) = {\mathfrak A}(\cW)$
and consequently the algebra ${\mathfrak A}_\kappa(\cW)$ is
generated by the operators 
$ \big(\alpha_{\lambda_\cW}(A)\big){}_{\Lambda_\cW Q_\kappa \Lambda_\cW{}^{-1}}$, 
$A \in {\mathfrak A}(\cW_0)$. Now by Lemma~\ref{covariance} 
$$ U(\lambda) \, 
\big(\alpha_{\lambda_\cW}(A)\big){}_{\Lambda_\cW Q_\kappa \Lambda_\cW{}^{-1}}
\, U(\lambda)^{-1} = 
\big(\alpha_{\lambda \lambda_\cW}(A)\big){}_{\Lambda 
\Lambda_\cW Q_\kappa \Lambda_\cW{}^{-1} \Lambda^{-1}} \, , 
$$
and the operators on the right hand side of this equality are,
for $A \in {\mathfrak A}(\cW_0)$, the generators of the algebra
$ {\mathfrak A}_\kappa(\lambda \cW)$. Thus 
$\alpha_\lambda({\mathfrak A}_\kappa(\cW)) \subset 
 {\mathfrak A}_\kappa(\lambda \cW)$. 
Replacing in this inclusion $\lambda$ by $\lambda^{-1}$ 
and $\cW$ by $\lambda \cW$ and making use of the
fact that $\alpha_\lambda{}^{-1} = \alpha_{\lambda^{-1}}$, one obtains 
$ {\mathfrak A}_\kappa(\lambda \cW)  \subset 
\alpha_\lambda({\mathfrak A}_\kappa(\cW))$. Hence
$\alpha_\lambda({\mathfrak A}_\kappa(\cW)) = 
 {\mathfrak A}_\kappa(\lambda \cW)$, \ie the deformed
algebras satisfy the condition of covariance as well. 

\pagebreak
     Turning to the proof of locality, we first restrict attention 
to the wedge $\cW_0$. According to fact (iii) mentioned above, one has 
$Q_\kappa V_+ = \cW_0$; hence 
$\cW_0 + Q_\kappa p \subset \cW_0$ and consequently  
$\cW_0^{\, \prime} \subset (\cW_0 + Q_\kappa p)^\prime$ for $p \in V_+$.
Since $V_+$ is a cone, this implies 
$(\cW_0^{\, \prime} - Q_\kappa q) \subset (\cW_0 + Q_\kappa p)^\prime$ 
for all $p,q \in V_+$. It then follows from the covariance and 
locality properties of the original algebras that for any
pair of operators $A \in {\mathfrak A}(\cW_0)$
and $B \in {\mathfrak A}(\cW_0^{\, \prime})$ one has 
(denoting the pure translations $(1,x) \in \cP$ by $x$)
$$ [\alpha_{Q_\kappa p}(A), \alpha_{-Q_\kappa q}(B)] = 0 \, ,
\quad p,q \in V_+ \, .$$
According to Lemma \ref{commutation}, this implies 
$[A_{Q_\kappa}, B_{-Q\kappa}] = 0$. Now if 
$\lambda = (\Lambda,x)$ is any Poincar\'e transformation 
such that $\lambda \cW_0 = \cW_0^{\, \prime}$, it follows from fact 
(ii) mentioned above that $\Lambda Q_\kappa \Lambda^{-1} = - Q_\kappa$. 
Hence the operators $B_{-Q\kappa}$, 
$B \in {\mathfrak A}(\cW_0^{\, \prime})$, generate the algebra 
${\mathfrak A}_\kappa(\cW_0^{\, \prime})$, and similarly the operators
$A_{Q_\kappa}$, $A \in {\mathfrak A}(\cW_0)$, generate
the algebra ${\mathfrak A}_\kappa(\cW_0)$. So we obtain the inclusion
${\mathfrak A}_\kappa(\cW_0^{\, \prime}) \subset 
{\mathfrak A}_\kappa(\cW_0){}^\prime$. By the Poincar\'e
covariance of the deformed algebras, 
established in the preceding step, it is then 
clear that  ${\mathfrak A}_\kappa(\cW^{\, \prime}) \subset 
{\mathfrak A}_\kappa(\cW){}^\prime$ for all $\cW \in \bcW$.

     It remains to establish the Reeh--Schlieder property of the
deformed algebras. According to Lemma \ref{leftright}, one has 
$A_Q = {}_QA$ for any skew symmetric matrix $Q$.  Hence 
$A_Q \Omega = {}_QA \Omega = \int \alpha_{Qp}(A) dE(p) \Omega = A \Omega$, 
since $\Omega$ is invariant under spacetime translations.  Thus 
${\mathfrak A}_\kappa(\cW) \, \Omega \supset {\mathfrak A}(\cW) \, \Omega$ 
for any $\cW \in \bcW$, so the Reeh--Schlieder property of the deformed wedge
algebras is inherited from the original algebras.  We summarize these
findings in a theorem.

\begin{theorem}
Let ${\mathfrak A}(\cW) \subset {\mathfrak F}$, $\cW \in \bcW$, be 
a family of wedge algebras having the Reeh--Schlieder 
property and satisfying the conditions of isotony, covariance, and 
locality. Then the family of deformed algebras 
${\mathfrak A}_\kappa(\cW) \subset {\mathfrak F}$, $\cW \in \bcW$,
introduced in Definition~\ref{def} also has these properties. 
\end{theorem}

     This theorem establishes that the deformation procedure outlined
above can be applied to any quantum field theory. If one 
starts with the wedge algebras in a free field theory, one arrives at 
the deformed theories considered in \cite{GrLe}, as can be
seen by explicit computations. But one may equally well take
as a starting point any rigorously constructed model, such as the 
self--interacting ${\cal P}(\varphi)$--theories in $d=2$ dimensions or the
$\varphi^4$--theory in $d=3$ dimensions \cite{GlJa}. 
In all of these cases, the
warped convolution produces a true deformation of the underlying 
theory, in the sense that the scattering matrix changes. 

     To exhibit this fact, let us assume that the underlying theory describes 
a single scalar massive particle. Then the spectrum of $U\rest \RR^d$ has 
the form 
$$
\mbox{sp} \, U\rest \RR^d 
= \{ 0 \} \cup \{ p : p_0 = \sqrt{\bcp^2 + m^2} \}
\cup \{ p : p_0 \geq \sqrt{\bcp^2 + M^2} \},
$$
with $M > m > 0$. In the present
general setting of wedge--local operators one can then define 
two--particle scattering states as in Haag--Ruelle--Hepp
scattering theory \cite{BoBuSch}. To see this, we fix the
standard wedge $\cW_0$ and pick operators $A \in {\mathfrak A}(\cW_0)$
and $A^\prime \in {\mathfrak A}(\cW_0{}^\prime)$
which interpolate between the vacuum vector $\Omega$
and single particle states of mass $m$. We then proceed to the deformed
operators $A_{Q_\kappa} \in {\mathfrak A}_\kappa(\cW_0)$, 
$A^\prime{}_{- Q_\kappa} \in {\mathfrak A}_\kappa(\cW_0{}^\prime)$
and note that these operators have the same interpolation properties
as the original ones, recalling that $A_{Q_\kappa} \Omega = A \Omega$, 
$A^\prime{}_{- Q_\kappa} \Omega = A^\prime{} \Omega$. 

     Next, we pick test functions $f, f^\prime \in \cS(\RR^d)$ whose
Fourier transforms $\widetilde{f}, \widetilde{f^\prime}$ have compact
supports in small neighborhoods of points on the isolated mass shell
in $\mbox{sp} \, U\rest \RR^d $ which do not intersect with the rest
of the spectrum.  With the help of these functions and the above
operators we define
$$
A_{Q_\kappa}(f_t) \doteq \int \! dx \, f_t(x) \, \alpha_x(A_{Q_\kappa}) \, ,
$$
where the functions $f_t \in \cS(\RR^d)$, $t \in \RR$, are given by 
\begin{equation} \label{hepp}
x \mapsto f_t (x) = (2\pi)^{-d/2} \! \! \int \! dp \, \widetilde{f}(p)
\, e^{i (p_0 - \omega_\sbcp)t} \, e^{-ipx} 
\end{equation}
with $\omega_\sbcp = (\bcp^2 + m^2)^{1/2}$. Similarly, 
one defines the operators $A^\prime{}_{-Q_\kappa}(f^\prime_t)$. 
Bearing in mind the support properties of $\widetilde{f},
\widetilde{f^\prime}$ and the preceding remark about the
action of the deformed operators on the vacuum vector, it follows that   
$A_{Q_\kappa}(f_t) \Omega = A(f_0) \Omega$ and
$A^\prime{}_{-Q_\kappa}(f^\prime_t) \Omega = 
A^\prime (f^\prime_0) \Omega$  are single particle states 
which do not depend on $t$. 

     The operators $A_{Q_\kappa}(f_t)$,
$A^\prime{}_{-Q_\kappa}(f^\prime_t)$ can be used to construct
incoming, respectively outgoing, two--particle scattering states. Yet
in the present case of wedge--localized operators this construction
requires a proper adjustment of the support properties of the Fourier
transforms of $f, f^\prime$. Introducing the notation $\Gamma(g)
\doteq \{ (1, \bcp/\omega_\sbcp) : p \in \mbox{supp} \, \widetilde{g}
\}$ for the velocity support of a test function $g$ and writing $g_1
\succ g_2$ whenever the set $\Gamma(g_1) - \Gamma(g_2)$ is contained
in the interior of the wedge $\cW_0$, one relies on the following
facts. According to a result of Hepp \cite{He}, the essential supports
of the functions $f_t$, $f_t^\prime$ are, for asymptotic $t$,
contained in $t \, \Gamma(f)$, $t \, \Gamma(f^\prime)$, respectively.
Moreover, the regions $\cW_0 + t \Gamma(f)$ and $\cW_0{}^\prime + t
\Gamma(f^\prime)$ are spacelike separated for $t<0$ ($t>0$) if
$f^\prime \succ f$ ($f \succ f^\prime$), respectively.  Because of the
covariance and locality properties of the deformed wedge--algebras,
one can then establish by standard arguments \cite{BoBuSch} the
existence of the strong limits
\begin{equation*}
\begin{split}
& \lim_{t \rightarrow -\infty} 
A_{Q_\kappa}(f_t) A^\prime{}_{-Q_\kappa}(f^\prime_t) \Omega
\doteq | A(f) \Omega \otimes_\kappa A^\prime(f^\prime) \Omega
\rangle^{\mbox{\scriptsize in}}  \quad \mbox{for} \ f^\prime \succ f \\
& \lim_{t \rightarrow \infty} 
A_{Q_\kappa}(f_t) A^\prime{}_{-Q_\kappa}(f^\prime_t) \Omega
\doteq | A(f) \Omega \otimes_\kappa A^\prime(f^\prime) \Omega
\rangle^{\mbox{\scriptsize out}} \quad \mbox{for} \ f \succ f^\prime \, .
\end{split}
\end{equation*}
The limit vectors have all properties of a symmetric tensor
product of the single particle states $A(f) \Omega$,
$A^\prime(f^\prime) \Omega$. In particular, they do not depend on the
specific choice of operators $A, A^\prime$ and test functions
$f, f^\prime$ within the above limitations. Because of the 
Reeh--Schlieder property of the wedge algebras, it is also clear
that these vectors form a basis in the respective asymptotic 
two--particle spaces.  

\pagebreak
     In order to exhibit the dependence of the tensor products on the
deformation parameter $\kappa$, we note that for $f^\prime \succ f$
\begin{equation*}
\begin{split}
& | A(f) \Omega \otimes_\kappa A^\prime(f^\prime) 
\Omega \rangle^{\mbox{\scriptsize in}}
= \lim_{t \rightarrow -\infty} 
A_{Q_\kappa}(f_t) A^\prime{}_{-Q_\kappa}(f^\prime_t) \Omega \\
& =  \lim_{t \rightarrow -\infty} 
\int \! dE(p) \, \alpha_{Q_\kappa p}(A)(f_t) \, 
A^\prime(f^\prime_t) \Omega \\
& =
\int \! dE(p) \, | U(Q_\kappa p) A(f) \Omega \otimes
A^\prime(f^\prime) \Omega \rangle^{\mbox{\scriptsize in}} \, ,
\end{split}
\end{equation*}
where the third equality follows from the fact that the limit can be
pulled under the integral and the symbol $\otimes$ denotes the tensor
product in the original theory. Similarly, one obtains for 
$f \succ f^\prime$
$$
| A(f) \Omega \otimes_\kappa A^\prime(f^\prime) \Omega
\rangle^{\mbox{\scriptsize out}} 
= 
\int \! dE(p) \, | U(Q_\kappa p) A(f) \Omega \otimes
A^\prime(f^\prime) \Omega \rangle^{\mbox{\scriptsize out}} \, .
$$
These relations between the scattering states in the original and 
in the deformed theory become more transparent if one proceeds 
to improper single particle 
states of sharp momentum $p = (\sqrt{\bcp^2 + m^2}, \bcp)$,
$q = (\sqrt{\bcq^2 + m^2}, \bcq)$. 
There one has 
\begin{equation*}
\begin{split}
|p \otimes_\kappa q \rangle^{\mbox{\scriptsize in}}
& = e^{i |p Q_\kappa q |} \, 
|p \otimes q \rangle^{\mbox{\scriptsize in}} \\
|p \otimes_\kappa q \rangle^{\mbox{\scriptsize out}}
& = e^{- i |p Q_\kappa q |} \, 
|p \otimes q \rangle^{\mbox{\scriptsize out}} \, .
\end{split}
\end{equation*}
The scattering states in the deformed theory depend on the matrix
$Q_\kappa$ through the choice of the wedge $\cW_0$ and thus break the
Lorentz symmetry in $d>2$ dimensions.  This can be understood if one
interprets the wedge--local operators as members of a theory on
non--commutative Minkowski space, where the Lorentz symmetry is broken
\cite{GrLe}.

     The kernels of the elastic scattering matrices in the 
deformed and undeformed theory are related by 
$${}^{out}{\langle p \otimes_\kappa q |p^\prime \otimes_\kappa q^\prime
  \rangle}{}^{in}
= e^{i|pQ_\kappa q| + i|p^\prime Q_\kappa q^\prime|} \; \; 
{}^{out} \langle p \otimes q | p^\prime \otimes q^\prime \rangle{}^{in} \, .
$$
Thus they differ from each other, showing that the 
deformed and undeformed theories are not isomorphic. Yet
since the difference is only a phase factor, the collision
cross sections do not change under these deformations. 
Hence the effects of the deformation, such as the asymptotic
breakdown of Lorentz invariance, could only be seen in
certain specific arrangements such as time delay experiments.

\section{Concluding remarks} 

     In the present article we have presented a generalization of
the deformation procedure of free quantum field theories, established 
by Grosse and Lechner \cite{GrLe}, to the general
setting of relativistic quantum field theory. Even though the new theories 
which emerge in this way may not be of direct physical relevance,
the results are of methodical interest. For they reveal yet again
the significance of the wedge algebra in the algebraic 
approach to the construction of models. 

     From the algebraic point of view the problem of constructing a
quantum field theory presents itself as follows. Given the stable
particle content in the situation to be described, one first
constructs a corresponding Fock space and representation $U$ of the
Poincar\'e group $\cP$.  A theory with this particle content is then
obtained by fixing a wedge $\cW_0$, say, and exhibiting a *--algebra
${\mathfrak G} \subset {\mathfrak F}$ which can be interpreted as the
algebra generated by fields and observables localized in
$\cW_0$. It thus has to satisfy the conditions
\begin{enumerate}
\item[(a)] $\alpha_\lambda({\mathfrak G}) \subset {\mathfrak G}$
whenever $\lambda \cW_0 \subset \cW_0$ for $\lambda \in \cP$.
\item[(b)]  $\alpha_{\lambda^\prime}({\mathfrak G}) \subset {\mathfrak G}^\prime$
whenever $\lambda^\prime \, \cW_0 \subset \cW_0^\prime$ 
for $\lambda^\prime \in \cP$.
\end{enumerate}
Any algebra ${\mathfrak G}$ satisfying these conditions  
is the germ of a quantum field theory in the following sense: setting
$ {\mathfrak A}(\cW) \doteq \alpha_\lambda({\mathfrak G})$,  
where $\lambda \in \cP$ is such 
that $\cW = \lambda \cW_0$ for given $\cW \in \bcW$,
it is an immediate consequence of the assumed properties
of $\mathfrak G$ that the definition of the
wedge algebras ${\mathfrak A}(\cW)$
is consistent and satisfies the conditions of
isotony, covariance and locality. As explained above, the
algebras corresponding to arbitrary causally closed regions
can then consistently be defined by taking intersections 
of wedge algebras. Conversely, any
asymptotically complete quantum field theory with the given
particle content fixes an algebra ${\mathfrak G}$ with
the above properties.
Thus any quantum field theory can in principle be presented 
in this way. However, at present a dynamical principle by which the
algebras ${\mathfrak G}$ can be selected is missing.

     Nevertheless, this algebraic approach has already proven to be
useful in the construction of interesting examples of quantum field
theories. For instance, the existence of an infinity of models in
$d=2$ spacetime dimensions with non--trivial scattering matrix was
established in this setting in \cite{Le,Le2,Le3}{}, thereby solving a
longstanding problem in the so--called form factor program of quantum
field theory, \cf \cite{Sch} and references quoted there. Wedge
algebras associated with a nonlocal field in $d \geq 2$ spacetime
dimensions were used in \cite{BuSu2} to construct local observables
manifesting non--trivial scattering.  Wedge algebras were also used in
\cite{BrGuLo2} for the construction of quantum field theories
describing massless particles with infinite spin, \cf also
\cite{MuSchYng} for a construction of operators in these theories with
somewhat better localization properties.

     The idea of deforming given wedge algebras in order to arrive at
new theories is a quite recent development in the algebraic approach
and sheds new light on the constructive problems in quantum field
theory. One may expect that the particular deformation procedure
considered here is only an example of a richer family of similar
constructions.  Moreover, these methods can also be transferred to
quantum field theories on curved spacetimes with a sufficiently big
isometry group.

     It is an intriguing question in this context to find manageable
criteria which allow one to decide whether the intersections of wedge
algebras are non--trivial. In \cite{BuLe} such a criterion based on
the modular structure was put forward.  Unfortunately, it is only
meaningful in $d=2$ spacetime dimensions.  In the examples of deformed
theories in $d > 2$ spacetime dimensions discussed here, it can be
shown that the algebras corresponding to bounded spacetime regions are
trivial.  But, viewing the deformed theory as living on
non--commutative Minkowski space \cite{GrLe}, one may expect that the
algebras corresponding to the intersection of two opposite wedges are
non--trivial. It would be of conceptual interest to establish this
conjecture.


\end{document}